**Reducing non-linear effects in Kelvin Probe Force Microscopy of back-gated 2D semiconductors**

Zander Scholl[1a], Ezra Frohlich[1a], Natalie Rogers[1a], Paul Nguyen[2], Baker Hase[2], Joseph Tatsuro Murphy[3], Joel Toledo-Urena[3], David Cobden[2], Jennifer T. Heath[1*]

[1]Department of Physics, Reed College, Portland, OR 97202 USA

[2]Department of Physics, University of Washington, Seattle, Washington 98195, USA

[3]Department of Physics, Linfield University, McMinnville, OR 97128, USA

*Correspondence to: jheath@reed.edu

In 2D field effect transistors the gate electrostatically dopes the 2D semiconductor (2DSC) channel, tuning the Fermi level. In principle, Kelvin probe force microscopy (KPFM) can detect the Fermi level, and its dependence on gate bias as well as position, potentially directly yielding band gaps, contact barriers, spatial nonuniformities, and sub-gap densities of states in such devices. However, KPFM relies on an oscillating probe voltage which itself electrostatically dopes the 2DSC, potentially creating a nonlinear response. Here, we show that when a suitably thin hBN back-gate dielectric is used, the KPFM signal agrees well with expectations, as explained by a quasistatic charge-balance model. Corresponding experimental results are consistent with the literature values of the bandgaps of monolayer and trilayer $WSe_2$. With this approach, the widely available technique of KPFM should find improved utility and new uses in the study of 2D devices.

Interest in 2D materials systems has exploded in recent years, as researchers have explored new materials and stacked them together in various combinations. Investigations range from manufacturable, room-temperature devices to systems exhibiting exotic physical behavior at low temperatures [1,2,3]. A tool that seems convenient and well suited for studying 2D materials and devices is Kelvin probe force microscopy (KPFM) [4], which in principle can map the electric potential variation with position or applied voltages.

In KPFM an AC voltage is applied between the metallic probe tip and the sample, creating an oscillating tip-sample force and force gradient which is detected via the tip's vibration amplitude (AM-KPFM) or resonance frequency shift (FM-KPFM) [5]. KPFM works well for metallic or semi-metallic surfaces, including graphene [6,7,8]. However, doped semiconductors require caution in interpreting the KPFM signal, $V_{\mathrm{KP}}$, as a direct quantitative measure of the Fermi level $E_{\mathrm{F}}$ due to the formation of a surface depletion region under the tip [9].

---

[a] Z.S., E.F., and N.R. contributed equally to this work.

While 2D semiconductors (2DSCs) have similarities to their bulk counterpart, they differ in that, by definition, they cannot support a surface depletion region that screens external fields. They instead respond by in-plane charge flow, which changes the charge density and can be described as "doping" the film. Here, we explore the conditions under which KPFM measurements can yield quantitative information about $E_F$ in a 2DSC.

KPFM has been employed on multiple occasions to study 2D field effect transistors (2DFETs). In current-voltage measurements of 2DFETs, contributions from contacts and spatial non-uniformity are hard to separate from the intrinsic response of the 2DSC channel. Tunneling current images provide spatial maps but are still hampered by nonideal contacts [10]. In this light, KPFM has been used to study contact barriers [11,12,13], characterize built-in potential and depletion regions at interfaces [14,15], measure relative changes in work functions with chemical doping [16], and identify features hidden in topographic scans [17].

KPFM can also be fruitfully applied to measure the change in $E_\mathrm{F}$ while varying the bias on a gate, $V_\mathrm{g}$, in a simple back-gate transistor geometry (Fig. 1a). In this way it has been used to investigate sub-bandgap defect densities [18], charge trapping in $MoS_2$ flakes [19], and gate-dependence of interfaces [20] and contact barriers [21].

Despite these apparent successes, there is reason to question the reliability of this KPFM data on 2DSCs. Given that the AC component of tip bias, $\tilde{V}_\mathrm{tip}$, is typically large enough to cause significant local band bending in the 2DSC, as shown schematically in Fig. 1b, the interplay of gate and tip voltages could distort the KPFM measurement. An important initial check is to observe $V_\mathrm{KP}$ when $E_\mathrm{F}$ is deep within the bandgap of the 2DSC. In this case there is no screening by free charge and $V_\mathrm{KP}$ should directly track $V_\mathrm{g}$. Reports of gating monolayer 2DSCs through $90\mathrm{nm}$ or $300\mathrm{nm}$ $SiO_2$ show a region of linear response, albeit with slopes $dV_\mathrm{KP}/dV_\mathrm{g}$ less than one [15,18,19]. Though their measurement conditions varied, these results suggest possible ambiguity in interpreting KPFM data.

For further useful deployment of KPFM in 2D materials systems there is an urgent need to eliminate this ambiguity. Here, therefore, we set out to carefully establish the conditions under which KPFM can reliably determine $E_\mathrm{F}$ in 2DSCs.

In KPFM, the tip is biased with a sum of DC and AC components,

$$V_\mathrm{tip} = V_\mathrm{DC} + \tilde{V}_\mathrm{tip}\sin(\omega_\mathrm{e} t). \quad (1)$$

This modulates the Coulomb interaction between the tip and sample, driving a mechanical tip oscillation at frequency $\omega_\mathrm{e}$. The value of $V_\mathrm{DC}$ is continuously adjusted in a feedback loop to minimize the resulting tip oscillations; the minimum should occur when the charge on the tip is zero [4]. $V_{\rm KP}$ is defined as this optimized value of $V_{\rm DC}$. For a metallic sample the zero-charge condition occurs at $V_{\rm DC} = V_0 = \Delta\varphi + V_{\rm S}$, where $\Delta\varphi$ equals the work function of the tip minus that of the sample, and $V_{\rm S}$ is the bias applied to the sample (typically, but not necessarily, zero). This condition assumes there is negligible potential offset due to fixed space charges in any dielectric separating the tip and the sample. The value $V_0$ is increased by decreasing the sample work function, or by applying a negative sample bias. Thus, an increase in $V_{\rm KP}$ corresponds directly to an increase in the sample's $E_{\rm F}$.

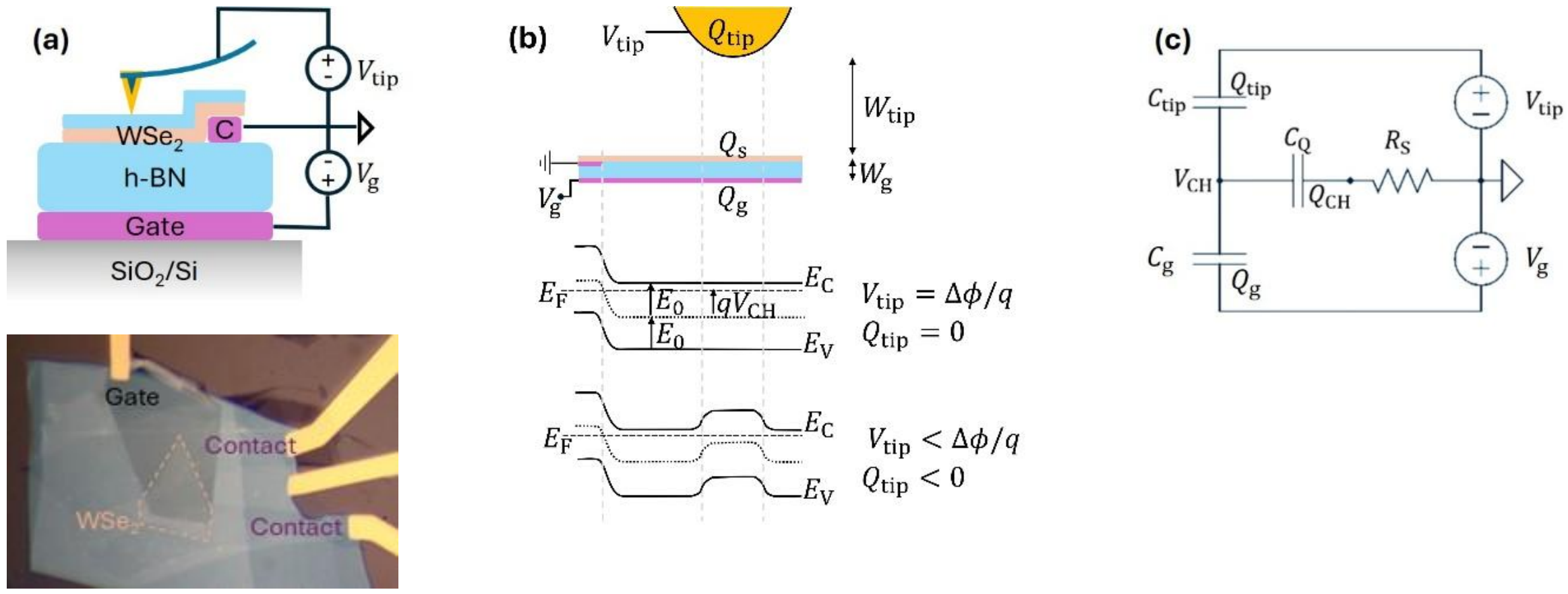

**Figure 1**. Experimental setup. **(a)** Above: schematic showing the typical sample stack with electrical connections to tip and gate voltages, $V_{\rm tip}$ and $V_{\rm g}$, where 'C' represents the grounded source/drain contacts to the $WSe_2$. Below: optical image of sample A. The monolayer $WSe_2$ channel is outlined for better visibility. **(b)** Schematic cross-section of the tip near the sample. The upper band diagram corresponds to the 2DSC with $V_{\rm g} > 0$, putting $E_{\rm F}$ near the conduction band. When the tip is biased to be neutral, it produces no band bending. The lower band diagram shows band bending near the tip when $V_{\rm tip}$ is more negative. In this geometry, at the contact, the gate dopes the underlying graphite flake, which carries current to the gold pads. **(c)** Schematic circuit diagram relating charges on the tip, gate, and channel ($Q_{\rm tip}$, $Q_{\rm g}$, and $Q_{\rm CH}$ respectively) to $V_{\rm tip}$ and $V_{\rm g}$ voltages and the channel potential $V_{\rm CH}$. The quantum capacitance $C_{\rm Q}$ embodies the nonlinear charge-potential relation of the channel (see text).

In the simple AM-KPFM technique, the tip motion at $\omega_{\rm e}$ is minimized. Other technique variants minimize the tip motion at sum or difference frequencies (sideband, heterodyne), or detect the resonance frequency modulation resulting from the tip motion (FM-KPFM) [22,23]. In these variants, the detected signal remains proportional to $V_{\rm DC} - V_0$, so our general discussion continues to be relevant, but the signal falls off more rapidly with distance, minimizing contributions from stray fields and allowing $V_0$ to be determined more accurately [24,25]. With these techniques, KPFM can be used to map spatial variations in $E_F$, such as those induced by gate doping of graphene [26]. See the supplemental materials and Fig. S4 for more discussion.

The goal is to relate $V_{\mathrm{KP}}$ to the local value of $E_{\mathrm{F}}$ in the 2DSC under the tip. A full three-dimensional self-consistent model would include factors such as tip shape and spatially resolved tip-sample interactions, as well as a complete 3D band model [27,28]. For a simple but practical approximation, we consider only a planar geometry, neglecting lateral variation in the sample and tip curvature, and take the charge response of the sample to be instantaneous, i.e., we assume electrostatic equilibrium under the tip. Our results below justify these assumptions.

In this model, the charges on the tip, $Q_{\mathrm{tip}}$, and gate, $Q_{\mathrm{g}}$, determine the charge on the sample, $Q_{\mathrm{tip}} + Q_{\mathrm{g}} = -Q_{\mathrm{S}}$, which is linked to $E_{\mathrm{F}}$ in the 2DSC. In accord with standard practice, we define the channel potential $V_{\mathrm{CH}}$ as the value of $E_{\mathrm{F}}$ relative to the midgap energy [29]; thus $E_{\mathrm{C}} - E_{\mathrm{F}} = E_0 - qV_{\mathrm{CH}}$ and $E_{\mathrm{F}} - E_{\mathrm{V}} = E_0 + qV_{\mathrm{CH}}$, where $E_{\mathrm{C}}$ and $E_{\mathrm{V}}$ are the conduction and valence band energies, respectively, $E_0 = (E_{\mathrm{C}} - E_{\mathrm{V}})/2$, and $q$ is the elementary charge, as shown in Fig. 1b. Tip and gate capacitances (per area), $C_{\mathrm{tip}}$ and $C_{\mathrm{g}}$ respectively, were estimated with a planar model using measured $W_{\mathrm{g}}$, $W_{\mathrm{tip}}$, and hBN cap thickness, and an hBN dielectric constant of 3.9. The schematic also includes the quantum capacitance $C_{\mathrm{Q}} = -\partial Q_{\mathrm{S}}/\partial V_{\mathrm{CH}}$ to conceptually illustrate the relationship between $Q_{\mathrm{S}}$ and $V_{\mathrm{CH}}$ . It should be noted, however, that $C_{\mathrm{Q}}$ is an imaginary, nonlinear capacitance. We must calculate the relationship between $Q_{\mathrm{S}}$ and $V_{\mathrm{CH}}$ directly within the model.

The electron and hole densities are given by

$$n = \int_{E_{\mathrm{C}}}^{\infty} g_{\mathrm{2D}}^{e} f(E)\,\mathrm{d}E = g_{\mathrm{2D}}^{e} kT \ln\{1 + \exp[-(E_0 - qV_{\mathrm{CH}})/kT]\} \quad (2a)$$

$$p = \int_{-\infty}^{E_{\mathrm{V}}} g_{\mathrm{2D}}^{h} f(E)\,\mathrm{d}E = g_{\mathrm{2D}}^{h} kT \ln\{1 + \exp[-(E_0 + qV_{\mathrm{CH}})/kT]\} \quad (2b)$$

with Boltzmann constant $k$, temperature $T$ and Fermi-Dirac distribution $f(E)$. Here, $g_{\mathrm{2D}}^{e}$ $\left(g_{\mathrm{2D}}^{h}\right)$ is the constant 2D electron (hole) density of states, $\pi g_{\mathrm{s}} g_{\mathrm{v}} m_{\mathrm{e}}^{*}/\hbar^{2}$, where $g_{\mathrm{s}}$ and $g_{\mathrm{v}}$ are the spin and valley degeneracies and $m_{\mathrm{e}}^{*}$ ($m_{\mathrm{h}}^{*}$) is the electron (hole) effective mass; we used $g_{\mathrm{s}} = g_{\mathrm{v}} = 2$ and $m_{\mathrm{e}}^{*} = m_{\mathrm{h}}^{*} = 0.4m_{\mathrm{e}}$ [30].

Balancing the charges then gives

$$C_T\left[V_{\mathrm{tip}} - V_{\mathrm{CH}}\right] + C_{\mathrm{g}}\left[V_{\mathrm{g}} - V_{\mathrm{CH}}\right] = q[p - n]. \quad (3)$$

To begin with, we took $Q_{\mathrm{tip}} = 0$, representing an ideal KPFM measurement, and solved Eq. 3 numerically for $V_{\mathrm{CH}}$ as a function of $V_{\mathrm{g}}$.

The model described thus far predicts the idealized behavior of a gated 2DSC, that one hopes to measure with KPFM. To investigate the nonlinear response of the 2DSC to $V_{\mathrm{tip}}$ during the KPFM measurement, the full expression of Eq. 3 was used with the first term nonzero. The value of $V_{\mathrm{CH}}$ was resolved at 100 points through the AC cycle, and averaged to yield the DC Fourier component $\langle V_{\mathrm{CH}}\rangle$; this process was repeated while iterating $V_{\mathrm{DC}}$ until $\langle V_{\mathrm{CH}}\rangle = V_{\mathrm{DC}}$. The final value is the modeled output $V_{\mathrm{KP}}$. As expected, the bands can be bent significantly by $V_{\mathrm{tip}}$, and the resulting asymmetry introduces an additional DC component that skews $V_{\mathrm{KP}}$, as can be seen in Fig. 2.

The preceding analysis relates the voltages on the tip, gate, and channel, modelling the KPFM response for a fairly general system. To quickly estimate the response of $V_{\mathrm{CH}}$ to *small* changes $dV_{\mathrm{tip}}$, one can use the circuit model illustrated in Fig. 1c, yielding

$$dV_{\mathrm{CH}} = \left(1 + \frac{C_{\mathrm{Q}}+C_{\mathrm{g}}}{C_{\mathrm{tip}}}\right)^{-1} dV_{\mathrm{tip}} \,. \tag{4}$$

This expression does not give a simple proportionality between the time-variation in $V_{\mathrm{CH}}$ and $V_{\mathrm{tip}}$, because $C_{\mathrm{Q}}$ changes during the AC cycle. It also cannot predict the impact of nonlinear response on $V_{\mathrm{KP}}$. However, it clearly demonstrates that if $R = C_{\mathrm{g}}/C_{\mathrm{tip}} \gg 1$ then $dV_{\mathrm{CH}}$ is divided down accordingly, minimizing any asymmetrical response.

Fig. 2 compares the modeled channel response for varying $R = C_g/C_{\mathrm{tip}}$. The illustrated results use $V_{\mathrm{DC}} = V_0$, meaning that $V_{\mathrm{DC}}$ is the correct KPFM output signal for the modeled sample. Thus, we should find $\langle \Delta V\rangle = \langle V_{\mathrm{tip}} - V_{\mathrm{CH}}\rangle = 0$, or equivalently, $\langle V_{\mathrm{CH}}\rangle = V_0$. In Fig. 2(a,d), $R = 0.6$, consistent with a 40nm tip-sample air gap, $5\mathrm{nm}$ hBN cap, and a $285\mathrm{nm}$ gate oxide of $SiO_2$ or h-BN. When $V_{\mathrm{g}} = 0$, $E_{\mathrm{F}}$ of the undoped 2DSC is mid-gap, and $V_{\mathrm{CH}}$ responds symmetrically to $\tilde{V}_{\mathrm{tip}}$, yielding the appropriate $\langle \Delta V\rangle = 0$. However, when $V_{\mathrm{g}} \neq 0$ the response of $V_{\mathrm{CH}}$ is asymmetric. For $V_{\mathrm{g}} = 1\ V$, as shown (Fig. 2d), $\langle \Delta V\rangle = 0.42$ V. In a KPFM experiment, this $\langle \Delta V\rangle$ would feed back into the control loop, erroneously indicating that $V_{\mathrm{DC}} \neq V_0$. Fig. 2(e,f) repeats this simulation for larger $R$, consistent with reducing the gate oxide thickness to 20nm. As suggested by Eq. 4, this reduces the nonlinear response of $V_{\mathrm{CH}}$. For $R = 8.3, \langle \Delta V\rangle = 0.03$, and the KPFM signal will be closer to its ideal value. Overall, this model suggests that the magnitude of the KPFM signal will be systematically too low when measuring 2DSCs, but that the error will become negligible for large enough $R$ (Fig. 2d) or when $|V_{\mathrm{g}}|$ increases enough that $C_{\mathrm{Q}} \gg C_{\mathrm{tip}}$ throughout the ac cycle of $V_{\mathrm{tip}}$. Additional model results are included in the supplementary materials.

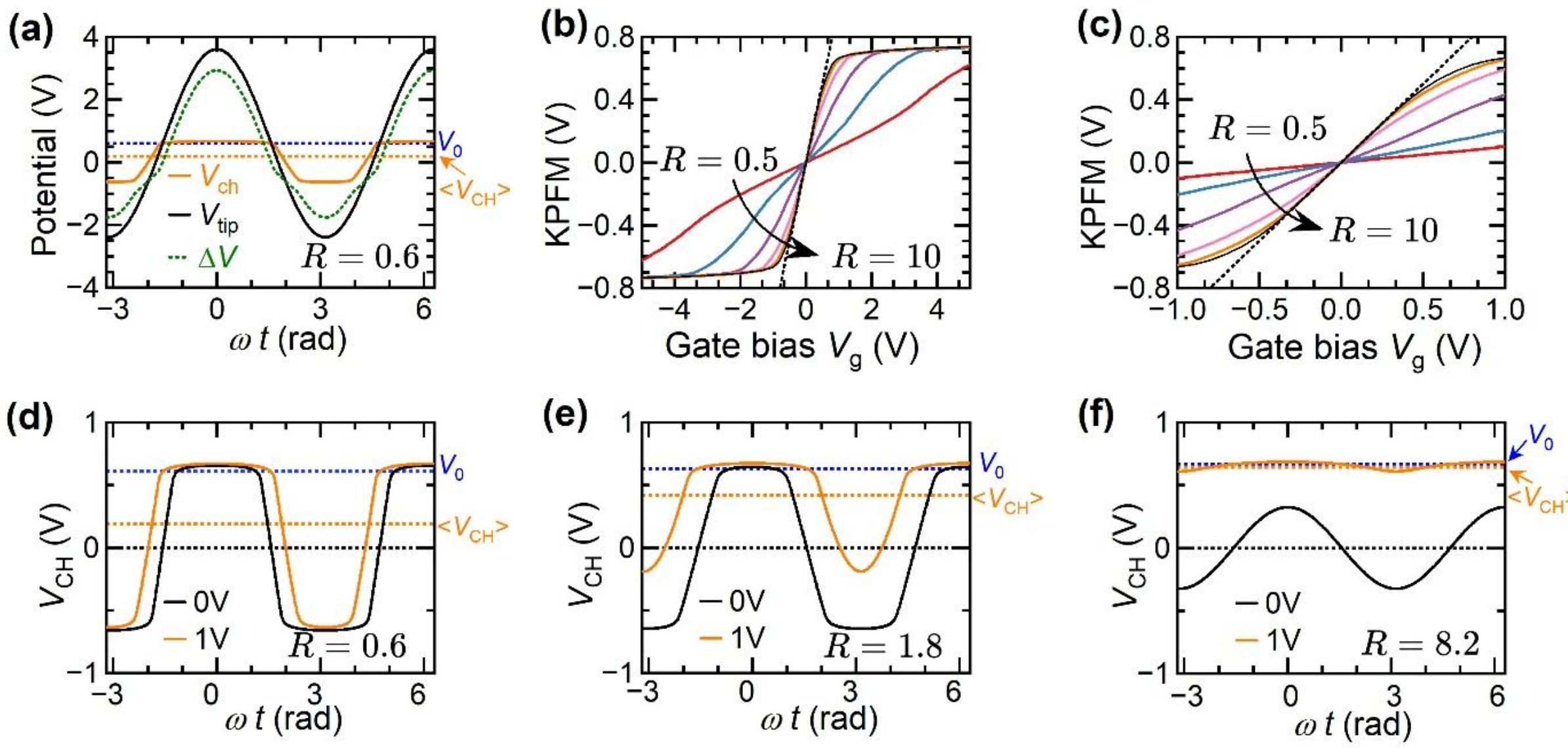

**Figure 2**. Simulations of the time-varying tip and channel voltage for a gated, undoped 2D semiconductor with $E_{\mathrm{g}} = 1.65$ V. In (a, d, e, f), $V_{\mathrm{tip}} = V_0 + 3V\cos(\omega_{\mathrm{e}} t)$ where $V_0$ is the channel potential ($V_{\mathrm{CH}}$) with no tip present. For an ideal KPFM measurement, then $\langle V_{\mathrm{CH}} \rangle = V_0$, such that $\langle \Delta V \rangle = \langle V_{\mathrm{tip}} - V_{\mathrm{CH}} \rangle = 0$. **(a)** Simulation using $R = 0.6$, consistent with $W_{\mathrm{g}} = 285$ nm, and $V_{\mathrm{g}} = 1$ V. Due to the nonlinear response of $V_{\mathrm{CH}}$ to $V_{\mathrm{tip}}$, $\langle V_{\mathrm{CH}} \rangle \neq V_0$. The KPFM feedback loop would thus adjust $V_{\mathrm{DC}}$, ultimately converging to an incorrect $V_{\mathrm{KP}}$. **(b)** Simulated $V_{\mathrm{KP}}$ for devices with $R = 0.5, 1, 2, 4, 7$, and 10 and varying $V_{\mathrm{g}}$. A line of slope 1 is included for reference (dashed). For larger $R$, $V_{\mathrm{KP}}$ more accurately measures $V_0$. $|V_{\mathrm{CH}}|$ increases asymptotically towards $E_g/2$ for larger $V_{\mathrm{g}}$ values. **(c)** Detail of (b). **(d-f)** Simulated $V_{\mathrm{CH}}$ for $V_{\mathrm{g}} = 0$ V and 1 V, and for varying $R$ as labeled, corresponding to (d) $W_{\mathrm{g}} = 285\mathrm{nm}$, (e) $W_{\mathrm{g}} = 90\mathrm{nm}$, and (f) $W_{\mathrm{g}} = 20\mathrm{nm}$. Values of $V_0$ and $\langle V_{\mathrm{CH}} \rangle$ are indicated for $V_{\mathrm{g}} = 1$ V (black and orange dotted lines respectively); when $V_{\mathrm{g}} = 0$ V, then $V_0 = \langle V_{\mathrm{CH}} \rangle = 0$ V (black dotted line). For thinner gate dielectrics (larger $R$), the nonlinear sample response becomes less significant.

As discussed above, several reports in the literature have this back-gated geometry and find $dV_{\mathrm{KP}}/dV_{\mathrm{g}} < 1$ in a low-doped region, possibly consistent with this model. However, the exact measurement and sample conditions are unknown, and indeed charge trapping at the $SiO_2$ interface means that $V_{\mathrm{g}}$ values are quite large in these reports. To further explore this response, we designed samples with both thick and thin gate oxide layers. In each sample, the multi-layer graphene gate electrode and the few-layer graphene contacts each connect to gold pads. The flakes were mechanically exfoliated, mainly using commercial crystals from 2D Semiconductors [31], and stacked using standard dry-polymer transfer techniques [32]. Thicknesses were measured with tapping mode atomic force microscopy. Three monolayer $WSe_2$ devices (A-C) have $W_{\mathrm{g}} = 22.0$, 13.5, and 18.5 nm respectively, $\pm 0.5$ nm. A fourth device (E) consists of 3-layer thick $WSe_2$ (3L-$WSe_2$) with $W_{\mathrm{g}} = 115 \pm 3$ nm.

Measurements were conducted with an Oxford Cypher S probe microscope using dual-pass FM-KPFM, under ambient conditions with the channel contacts electrically grounded. For each scan line, a first sweep measures the topography while tapping the tip into repulsion with amplitude $A$ corresponding to the average height of the tip above the hBN cap. In a second "nap" sweep, the tip follows the established topography, typically with height $W_{\mathrm{T}} = A$, but with the mechanical oscillation amplitude reduced to 10-20% of the initial value (~$10\ \mathrm{nm}$) so the tip-sample interaction is attractive, and the KPFM signal is measured. This procedure aims to determine $W_{\mathrm{T}}$ as accurately as possible. Tapping fully into mechanical contact wouldn't normally be advisable for KPFM; here, tip-sample electrical contact is avoided due to the hBN cap layer. We do not normally observe the contact electrification effect described in [33] as long as we avoid imaging over exposed $SiO_2$ regions.

During analysis, KPFM data were averaged using at least 100 pixels in each distinct region, avoiding large bubbles or other obvious defects. The pixel-wise standard deviations were $< 50\ \mathrm{mV}$ for $W_{\mathrm{T}} \leq 60\mathrm{nm}$, and $\sim 200\ \mathrm{mV}$ for $W_{\mathrm{T}} = 80\mathrm{nm}$. Additional scan to scan variation was generally $<\ 50\ \mathrm{mV}$, and could be due to stray charge, variations in the ambient surface, or changes in the tip between subsequent scans. Example KPFM images, taken with $\tilde{V}_{\mathrm{tip}} = 3\ \mathrm{V}$, are shown in Fig. 3, along with summative data. Additional sample data is included in the supplemental materials; sample response agreed well with expectations. For example, in the exposed hBN region, we expect $V_{\mathrm{KP}}$ to simply track the potential of the underlying graphite flake and thus the gate bias $V_{\mathrm{g}}$. All measured values of $dV_{\mathrm{KP}}/dV_{\mathrm{g}}$ were indeed in the range 0.93 to 1.06. The data can be modeled using a bandgap of 1.65 eV for samples A-C (monolayer $WSe_2$), and 1.1 - 1.2eV for sample E (3L-$WSe_2$), consistent with the literature [34,35,36].

Fig. 3 shows that for devices A-C, the data was reasonably described by the simple model with no tip-sample interaction, suggesting that the KPFM measurement works well. The model was improved by adding an exponentially distributed "tail" of electron acceptors near the conduction band edge [37]

$$n_{\mathrm{trap}} = a g_{\mathrm{oT}} \exp[-(E_0 - qV_{\mathrm{CH}})/a]. \qquad (5)$$

with $g_{\mathrm{oT}} = 1.4 \times 10^{13} \mathrm{cm}^{-2}\mathrm{eV}^{-1}$ and $a = 0.20\ \mathrm{eV}$, values consistent with prior measurements on $MoS_2$, where the defects were postulated to be S vacancies [18]. Such a defect band would slow the tuning of $E_{\mathrm{F}}$ into the upper half of the bandgap and immobilize much of the negative charge, degrading the transistor characteristics. Any shift in the contact work function with $V_{\mathrm{G}}$ was neglected; see the supplementary materials for data.

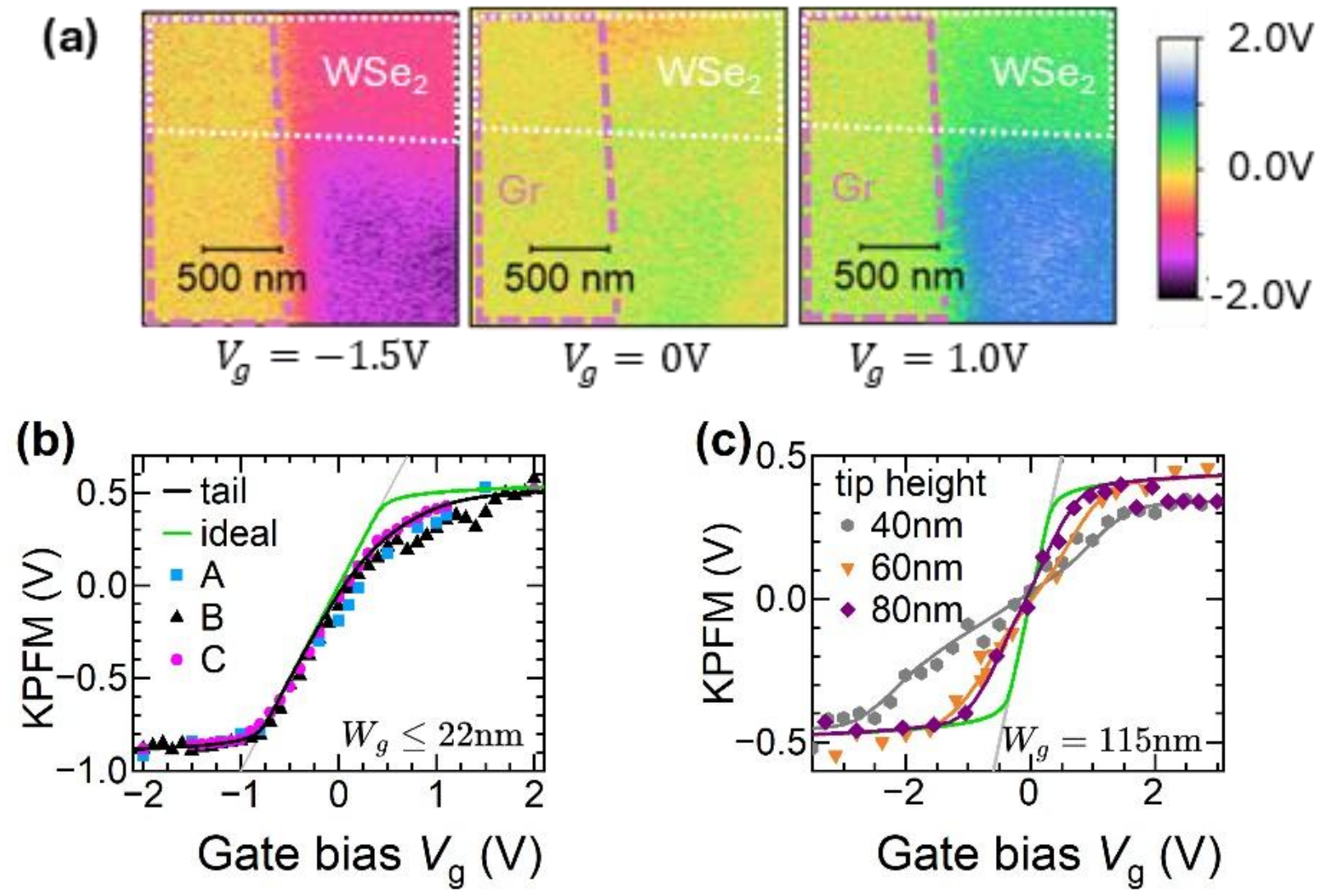

**Figure 3. (a)** KPFM maps of sample C for varying gate bias, showing the $WSe_2$ channel and its bilayer graphene contact, as well as, at the lower right of each image, the exposed hBN with the gate electrode underneath. **(b)** Three monolayer $WSe_2$ samples with thin hBN gate dielectrics, modeled using $E_g = 1.65$ eV. Here, $R \geq 8$ and the "ideal" model of Eq. 3 with $Q_{tip} = 0$ (green line) works well. Incorporating band tail states, Eq. 5, improves the model (black line). A line of slope 1 is included for reference. **(c)** Sample E, with 3l- $WSe_2$ has a thicker, $115$nm h-BN gate insulator. By varying $W_t$, $R$ was changed from 1.4 to about 4. The data are noisier than for samples A-C; the signal is weaker, and small amounts of charge impact the voltage more. Models use $E_g = 1.1$ eV (grey line) and $1.2$ eV (others). For additional detail, see the supplementary materials.

In contrast to devices with thin gate dielectrics, the response of device E, which has $W_g = 115$ nm, is not well modeled by Eq. 3 with $Q_{tip} = 0$. We hypothesize that nonlinear tip-sample interactions measurably influence the data for this geometry. Indeed, $V_{KP}$ is well described by the iterative model discussed above.

To better test the model, we measured sample E using $W_t$ of $40$, $60$, and $80$ nm. This was the extent that the tip height could be practically varied; lower heights cause the tip to interact with the sample, and with increased height the signal-to-noise degrades. A simple parallel-plate model using these $W_t$ values yields $R = 1.4, 2.1, 2.8$, respectively. However, one would expect the parallel-plate model to increasingly overestimate $C_t$ as $W_t$ increases, especially beyond the tip diameter of $40$ nm [24]. Indeed, models using $R = 1.4, 2.7, 4$ were in better agreement with data. These are the solid lines shown in Fig. 3c.

These model and experimental results show that, with appropriate sample geometry, $V_{CH}$ can be measured by KPFM even when there is very little charge on the channel, which is somewhat counterintuitive, but important for measuring interfaces and densities of states in working devices. It's easier to understand by considering that, in low doped samples, the condition $\langle V_{\mathrm{tip}} - V_{\mathrm{CH}} \rangle = 0$, which yields zero vertical electric field under the tip, remains meaningful even when the result is $V_{\mathrm{DC}} = \langle V_{\mathrm{CH}} \rangle = V_{\mathrm{g}}$.

We strive here for a simple model that yields intuitive understanding, and it does describe our devices well. Some important assumptions are (i) that the forces on the tip remain consistent with typical models [4,25], and (ii) that the ideal 2D approximation is sufficient to describe $g_{2\mathrm{D}}$. For example, in thicker samples the measured top-layer could differ from the gated bottom layer.

This work shows that KPFM measurements can be used productively to measure 2DSC devices. It would be hard to quantify data from ungated 2D samples, though, and gating through $SiO_2$ poses challenges. Instead, best results are obtained when $C_{\mathrm{g}} \gg C_{\mathrm{t}}$. Then, with a sweep of $V_{\mathrm{g}}$ one can assess parameters including the effectiveness of gating, sample bandgap, potential barriers at contacts, band-edge defects densities, and spatial nonuniformity. Any nonlinear tip-sample interaction can be identified from a mid-gap slope, $dV_{\mathrm{KP}}/dV_{\mathrm{g}} < 1$. For non-ideal samples, the KPFM signal will still be usable in regimes where $C_{\mathrm{Q}}$ remains large throughout the oscillation of $V_{\mathrm{tip}}$, and the nonlinear response at lower doping levels can be modeled with our simple iterative approach.

## Supplementary Material

See the supplementary material for more details on sample preparation and AFM methods, as well as additional KPFM measurement and model results.

## Author Declarations

The authors have no conflicts to disclose.

Primary data for this study are shared in the supplementary materials. Additional data that support the findings of this study are available from the corresponding author upon reasonable request.

## Acknowledgements

Thanks to Jay Ewing for assistance with design, machining, and fabrication, and to Keith Jones for guidance on the Cypher instrument and KPFM. J.H., J.M., and J.T.-U. were supported by the Murdock Charitable Trust Awards 202118943 and 2017319, as well as by the National Science Foundation (NSF) under DMR-MRI-1827971. N.R. was funded by the NSF DMR-2226592. Work at the University of Washington was supported by National Science Foundation (NSF) MRSEC award 2308979, and by Programmable Quantum Materials, an Energy Frontier Research Center funded by the U.S. Department of Energy (DOE), Office of Science, Basic Energy Sciences (BES), under award DE-SC0019443. The authors thank Dr. Ronen Dagan and Yossi Rosenwaks for sharing data and information.

**Supplementary Materials for: Reducing non-linear effects in Kelvin Probe Force Microscopy of back-gated 2D semiconductors**

Zander Scholl[1a], Ezra Frohlich[1a], Natalie Rogers[1a], Paul Nguyen[2], Baker Hase[2], Joseph Tatsuro Murphy[3], Joel Toledo-Urena[3], David Cobden[2], Jennifer T. Heath[1*]

[1]Department of Physics, Reed College, Portland, OR 97202 USA

[2]Department of Physics, University of Washington, Seattle, Washington 98195, USA

[3]Department of Physics, Linfield University, McMinnville, OR 97128, USA

*Correspondence to: jheath@reed.edu

## Sample preparation

Samples were prepared using standard mechanical exfoliation and dry stacking techniques [1]. Multilayer graphene (mL-Gr) contact layers were placed on top of evaporated gold pads. Film thicknesses were determined using tapping mode atomic force microscopy (AFM), while tapping into the repulsive regime, meaning that the phase signal was less than 90°. The z-piezo and z-sensor signals in the instrument agree with each other, and results are consistent with values that would be expected given the contrast seen in optical images. Thickness could be slightly over-estimated because data were collected in ambient conditions. Examples of raw data collected during the initial sample preparation process are shown in Fig. S1.

Similarly measured parameters for all samples are shown in Table S1. These consist of samples A-C, with relatively thin gate h-BN layers on the order of 20 nm, and sample E with a 115 nm h-BN layer. Sample E is more directly comparable to samples reported in the literature that were directly exfoliated onto 90 nm $SiO_2$. However, with the h-BN gate dielectric, samples do not have significant trapped charge at interfaces, and so can be measured without large offsets in $V_g$. For the KPFM measurement, this also reduces concerns about stray fields affecting the tip response.

Our sample “D” has been excluded from discussion here because it did not have an h-BN cap, and this was likely the reason that its data were significantly noisier than for the other samples. It was also measured early-on, and later measurements used a wider range of $V_g$. Results on sample D were otherwise entirely in agreement with data from samples A-C.

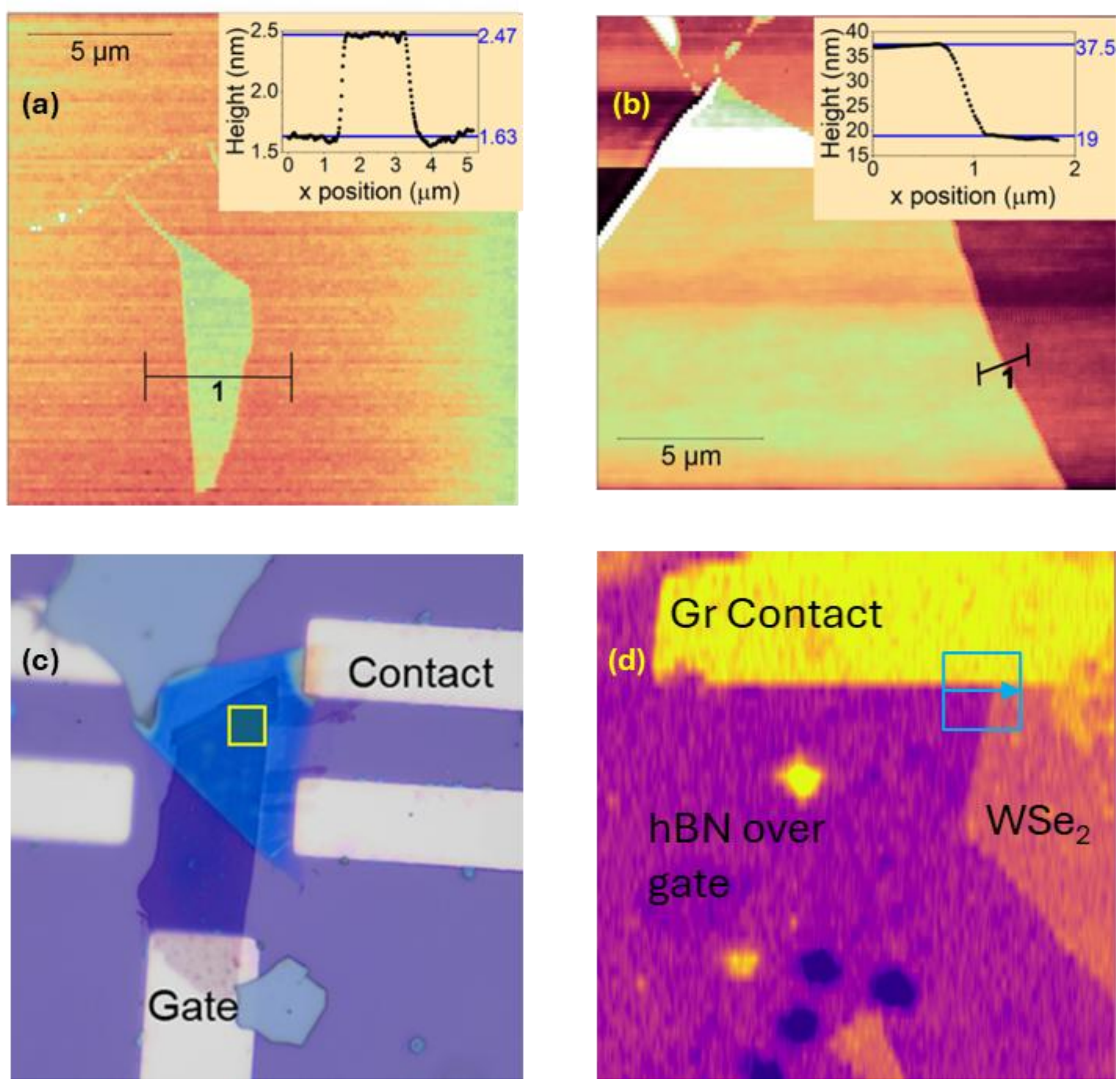

**Fig. S1**. Preparation of sample C. **(a)** Tapping mode AFM image of $WSe_2$, $0.8$ nm thick, consistent with a monolayer [2]. **(b)** Tapping mode AFM image of gate hBN, $18.5 \pm 0.5$ nm thick. **(c)** Optical image, stack of multilayer graphene (mL-Gr) and gate hBN over the mL-Gr gate contact. $WSe_2$ and capping hBN layers were added after this (but are nearly invisible in optical images). Yellow box indicates the approximate location of the KPFM image shown in d. **(d)** Example KPFM image collected with $V_{\mathrm{g}} = 1\mathrm{V}$. The KPFM data graphed in the text, Fig. 3, were obtained from a 1um region indicated by the blue box.

*Alt text: Atomic Force Microscopy and optical images showing sample details and assembly.*

**Table S1**. Key layer thicknesses for samples A-E. Uncertainties indicate standard deviations in the averaged area.

| Sample | h-BN gate oxide (nm) | $WSe_2$ channel (nm) | h-BN cap oxide (nm) |
|---|---|---|---|
| A | $22.0 \pm 0.5$ | $0.7 \pm 0.1$ | $3.0 \pm 0.5$ |
| B | $13.5 \pm 0.5$ | $0.8 \pm 0.1$ | $3.5 \pm 0.5$ |
| C | $18.5 \pm 0.5$ | $0.8 \pm 0.1$ | $5 \pm 0.5$ |
| E | $115 \pm 3$ | $2.0 \pm 0.5$ | $3.0 \pm 0.5$ |

*Alt text: Table showing gate oxide, channel, and cap layer thicknesses for each sample.*

### Atomic Force Microscopy calibration of cantilever optical lever

In preparation for dual pass FM-KPFM measurements, the instrument is calibrated following guidelines provided by Oxford Asylum Instruments. To determine lift height and tapping oscillation amplitudes for the KPFM imaging, the spring constant and optical lever (INVOLS) were found using the GetReal function. This is built into the instrument software, and uses the lever geometry combined with the thermal excitation data. For tapping-mode imaging, a standard "tune" was usually done, which finds the resonance frequency of the cantilever using mechanical (rather than thermal) excitation. Images of typical tuning and thermal curves are shown in Fig. S2.

After this calibration, the tip is engaged onto the sample an ac response-distance curve is collected. It is checked to ensure that the tip moves between attractive and repulsive interactions with the surface. If it is not possible to achieve a repulsive interaction with typical values of tip oscillation, then either the sample is cleaned or the tip is replaced. The tip is then moved to 300 nm above the sample surface for calibration of the electrical signal phase.

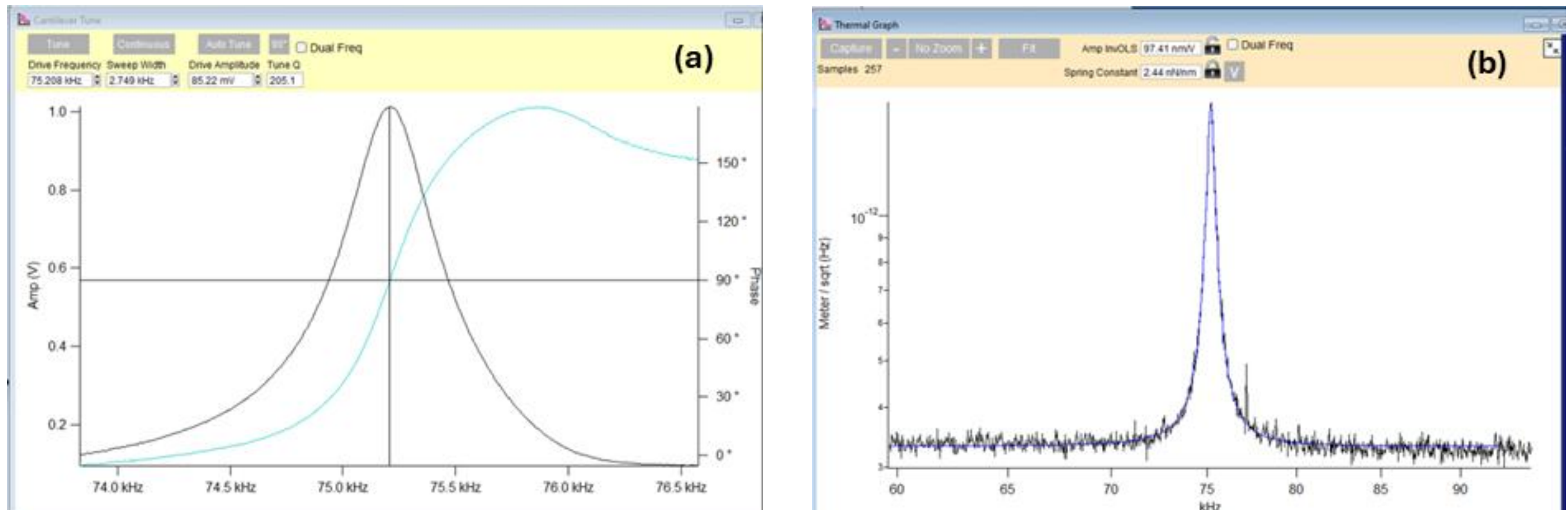

**Figure S2**. AFM calibration. (**a**) Typical tuning of an ASYELEC 01.R2 cantilever used for KPFM imaging, amplitude and phase, and **(b)** thermal calibration of the optical lever. Generally, stiffer AC160TSR3 cantilevers were used to measure sample thicknesses.

*Alt text: Output of the Cypher AFM software, demonstrating tuning of the tip resonance frequency.*

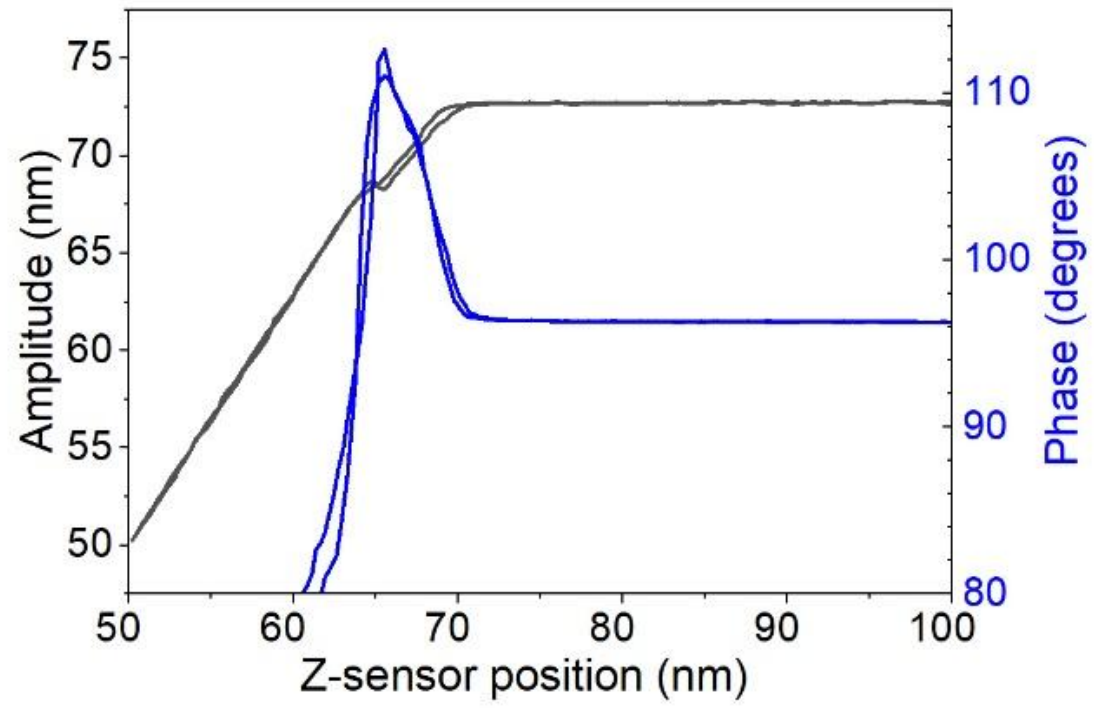

**Figure S3**. AC tip-sample approach curve using a tip oscillation amplitude of 100 nm, an approach setpoint of 50 nm, and with the sample contacts and gate grounded. The Z-sensor raw reading was offset to have a minimum value of 50nm in this graph, to make the relation between position and amplitude clearer. This shows that as the tip approaches the surface, it is initially attracted to the surface (phase increases $> 90°$ ), and then repelled. The 90° phase is tuned, as shown in Fig. S2, significantly above the surface, and is typically a few degrees higher when the tip is near the surface, even if the tip is not strongly interacting with the surface, as can be seen in the longer-distance data above.

*Alt text: Graph of tip oscillation amplitude and phase as a function of z-sensor position, showing both attractive and repulsive regimes of interaction.*

### KPFM theory

This will briefly outline some basic aspects of the theory of KPFM. More complete treatment can be found in [3].

As discussed in the main text, in KPFM, the tip is biased with a sum of DC and AC components,

$$V_{\mathrm{tip}} = V_{\mathrm{DC}} + \tilde{V}_{\mathrm{tip}} \sin(\omega_{\mathrm{e}} t). \tag{S1}$$

This modulates the Coulomb interaction between the tip and sample, driving a mechanical tip oscillation at frequency $\omega_{\mathrm{e}}$. The value of $V_{\mathrm{DC}}$ is adjusted in a feedback loop to continuously minimize the resulting tip oscillations. The reported KPFM signal $V_{\mathrm{KP}}$ is then equal to this optimized value of $V_{\mathrm{DC}}$. The logic behind this is as follows. The force $F_{\mathrm{e}}$ pulling the tip towards the sample is $\frac{\partial U}{\partial z}$, where $z$ is the tip height and $U$ is the energy of the tip-sample system. As $V_{\mathrm{tip}}$ is not a function of $z$, this becomes equivalent to

$$F_{\mathrm{e}} = \frac{1}{2}\left(V_{\mathrm{tip}} - V_0\right)\partial Q/\partial z, \tag{S2}$$

where $Q$ is the charge on the tip; this force disappears for $Q = 0$.

Because an electrochemical potential difference between the tip and sample will drive a current in the external circuit, as shown in Fig. S4, then for a metallic sample the zero charge condition occurs at $V_{\mathrm{tip}} = V_0 = \Delta E_{\mathrm{F}} + V_{\mathrm{S}}$. Here, $\Delta E_{\mathrm{F}}$ is related to the tip and sample work functions, $\varphi_{\mathrm{T}}$ and $\varphi_{\mathrm{s}}$ respectively, as $\Delta E_{\mathrm{F}} = \Delta\varphi = \varphi_{\mathrm{T}} - \varphi_{\mathrm{s}}$, and $V_{\mathrm{S}}$ is any bias applied to the sample (typically, but not necessarily, zero). The measured value can also be offset by any fixed space charges between the tip and sample, which can be viewed as shifting the value $\mathrm{V_S}$.

The total capacitance $C$ between tip and sample can then be *defined* according to

$$Q = C\left(V_{\mathrm{tip}} - V_0\right). \tag{S3}$$

As long as the sample is sufficiently metallic and the dielectric above it is linear, then $C$ is independent of $V_{\mathrm{tip}}$. From Eqs. (S1) to (S3) the component of $F_{\mathrm{e}}$ at $\omega_{\mathrm{e}}$ is then

$$F_{\mathrm{e},\omega}(z, V_0) = \frac{\partial C}{\partial z}\tilde{V}_{\mathrm{tip}}(V_{\mathrm{DC}} - V_0) \tag{S4}$$

This vanishes when $V_{\mathrm{DC}} = V_0$ and thus $V_{\mathrm{KP}}$ directly measures $\Delta\phi$. If the sample is gated, such that $E_{\mathrm{F}}$ increases, then by the same amount, $\varphi_{\mathrm{s}}$ decreases and $V_{\mathrm{KP}}$ increases. Thus, a change in the measured KPFM signal directly measures a change in $E_{\mathrm{F}}$ and $V_{CH}$.

While a simple AM-KPFM measurement, as described above, directly minimizes the tip motion due to $F_{\mathrm{e},\omega}$, other variants detect the tip motion at sum or difference frequencies (sideband or heterodyne, respectively), or detect the resulting resonance frequency modulation due to the tip motion (FM-KPFM) [4]. In these variants, the detected signal remains proportional to $V_{\mathrm{DC}} - V_0$, so the general discussion above continues to be relevant.

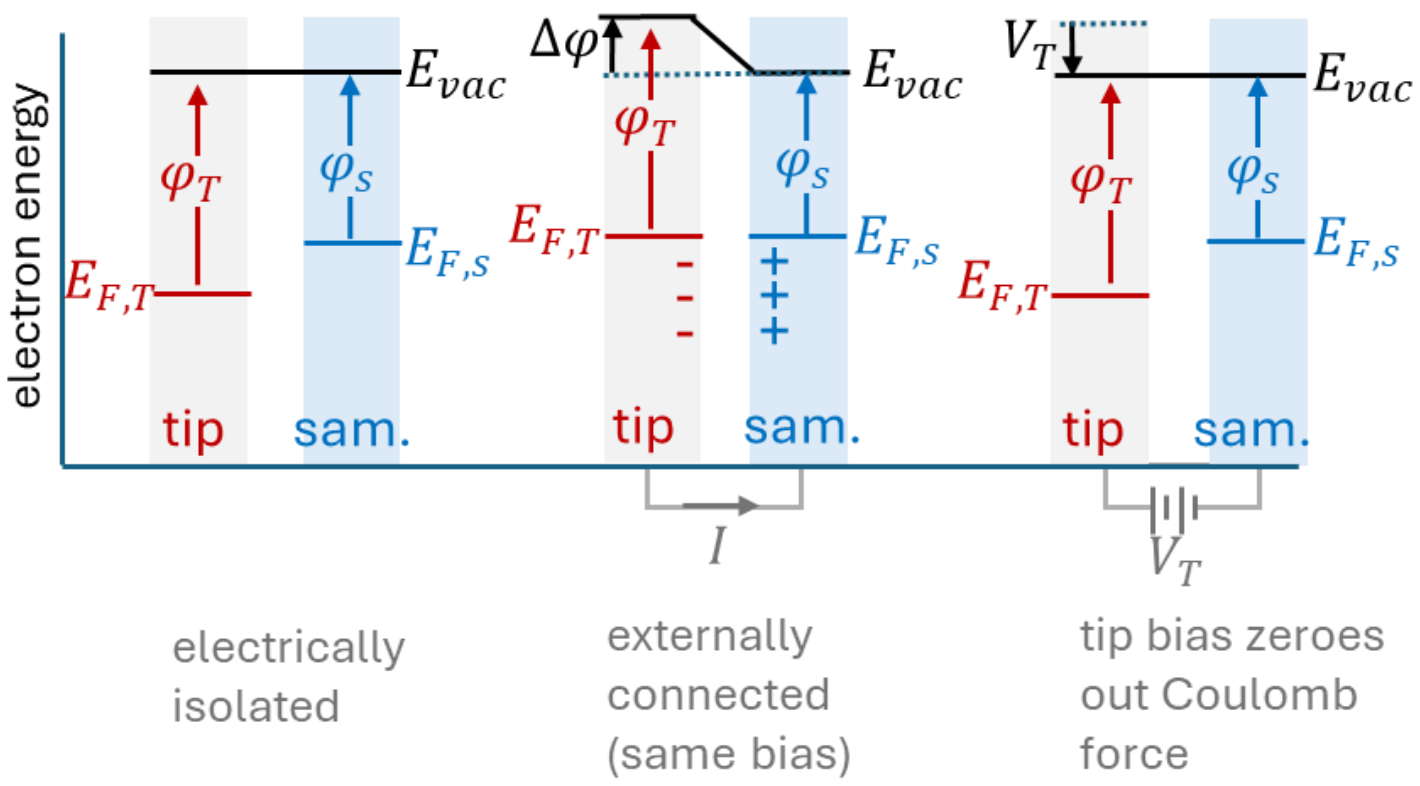

**Figure S4**. Visualization of the energies and biases in the tip-sample system. (left) tip and sample are electrically isolated and neutral. There is no electric field, and so the vacuum energy is flat. (center) When the tip and sample are shorted together, or equivalently, electrically connected externally to the same bias, electrons move from the higher value of $E_{\mathrm{F}}$ and onto the material with lower $E_{\mathrm{F}}$. Charge accumulates, and an electric field and electrical potential build up between the tip and sample. (right) The KPFM system adjusts $V_{\mathrm{T}}$ to zero out the charge and electric field between the tip and sample, which occurs when $V_{\mathrm{T}} = \Delta\varphi$. Any sample bias would add to this value of $V_{\mathrm{T}}$.

*Alt text: Energy diagram defining work function and showing the effect of tip bias.*

However, the signal from these variants is proportional to $\partial^2 C/\partial z^2$ and thus falls off more rapidly with distance. This minimizes contributions to the tip response from stray fields interacting with the sidewalls of the tip and the cantilever, allowing the value of $V_0$ to be determined more accurately [5,6]. For this reason, FM-KPFM was chosen for this study.

**KPFM results**

The different regions of the sample were known from optical microscopy images taken during the stacking procedure, and are more intuitively visible in the gated KPFM image as can be seen in Figure S5. During the topography scan, the tip ideally remained in repulsive mode, characterized by a phase <90°. During the "nap" scan when KPFM data is collected, the tip oscillation is reduced to about 20% of the initial value while normally the tip remains at the same average distance $W_{\mathrm{T}}$. This puts the tip in the attractive, or non-contact, regime,

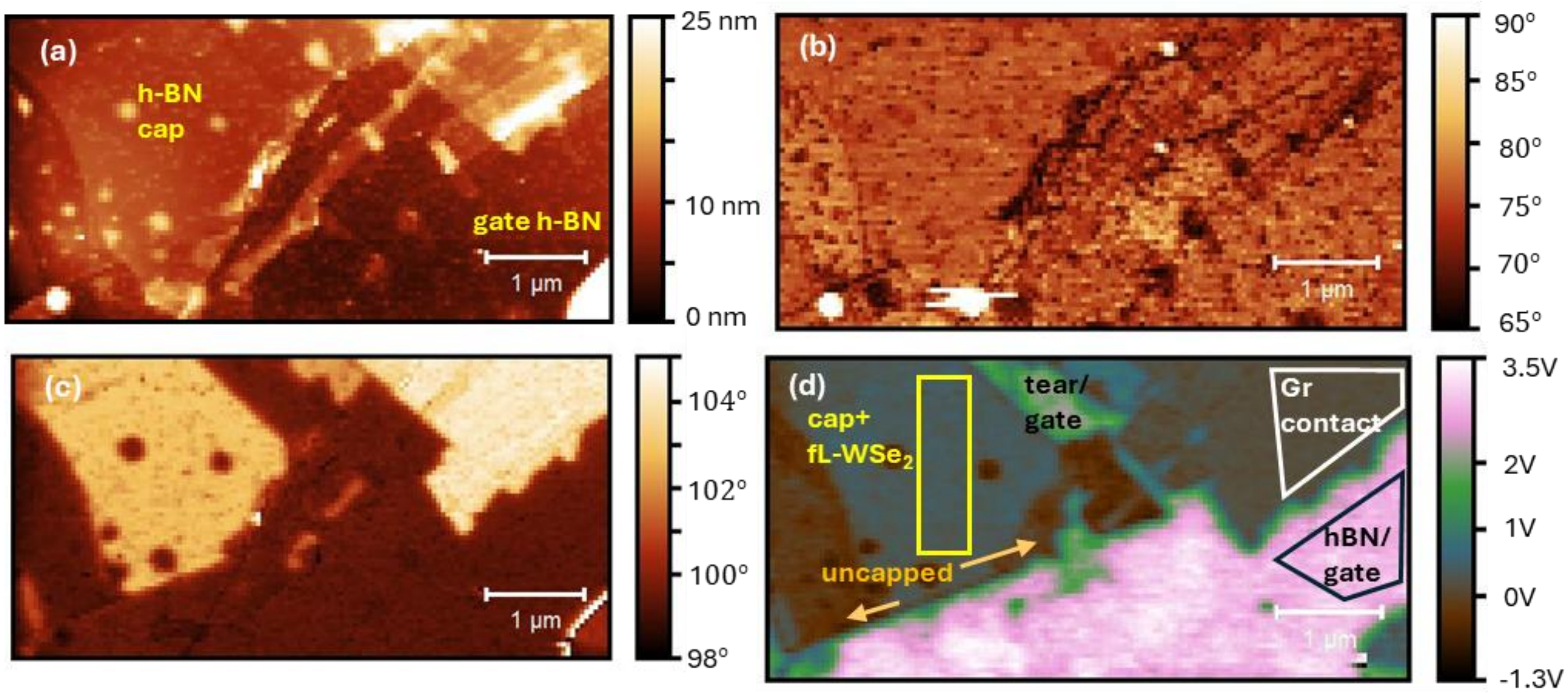

**Figure S5**. Measurement of sample E with contacts grounded and gate bias at 2.5V, including (**a**) Topography, dominated by the BN cap (which became somewhat fragmented when transferred; fortunately there is a large uniform cap over the important fL-$WSe_2$ region at upper left.) Prominent bubbles were avoided during data analysis. (**b**) Phase signal during the topography scan, in repulsive mode, illustrating phase <90° as desired, except in a few discrete locations with tall bubbles. (**c**) Phase signal during the nap scan while KPFM data were being collected, in attractive mode with phase >90°, and (**d**) FM-KPFM signal showing more clearly the functional regions of the sample. This sample was not beautiful, but it did the job. Uncapped areas and large bubbles were more negative. There is also a tear in the $WSe_2$ in the upper center region, as labeled. (the contact connected well with the $WSe_2$ flake in a region out of the image, above the tear). Regions selected for analysis are approximately indicated.

*Alt text: AFM and KPFM maps of sample E with measurement regions labeled.*

and the phase is >90°. We did try on some occasions to conduct the nap scan as close to the surface as possible. When it got too close, there would be glitches in the KPFM response, and in severe cases the feedback loop would invert, causing the signal to run away. So, if the tip were not in attractive mode during KPFM imaging, it was quite apparent.

Sometimes (for samples A-C) the phase during topography scanning was closer to 95°, but likely the tips were still in repulsive mode (see Fig. S3). Regardless, the response of these three samples was not sensitive to small variations in tip height and indeed the results for all three samples look very similar to each other. This can be seen in Fig. S6 and in the main paper. For sample E, the tip-sample distance was more critical and during KPFM measurements, the first pass topography scans were rigorously taken with phase <90°.

Figure S6 shows the raw data corresponding to results for samples A-C in the main paper. Here, we also include the data from the Gr contact and exposed h-BN regions, and the corresponding linear curvefits. The only difference from the results in the main paper is that in the paper, since all the plots are on the same graph, the $WSe_2$ data was shifted vertically to better align the data. Such shifts were: Sample A: -4.6mV, Sample B: +6.5 mV, Sample C: -100mV.

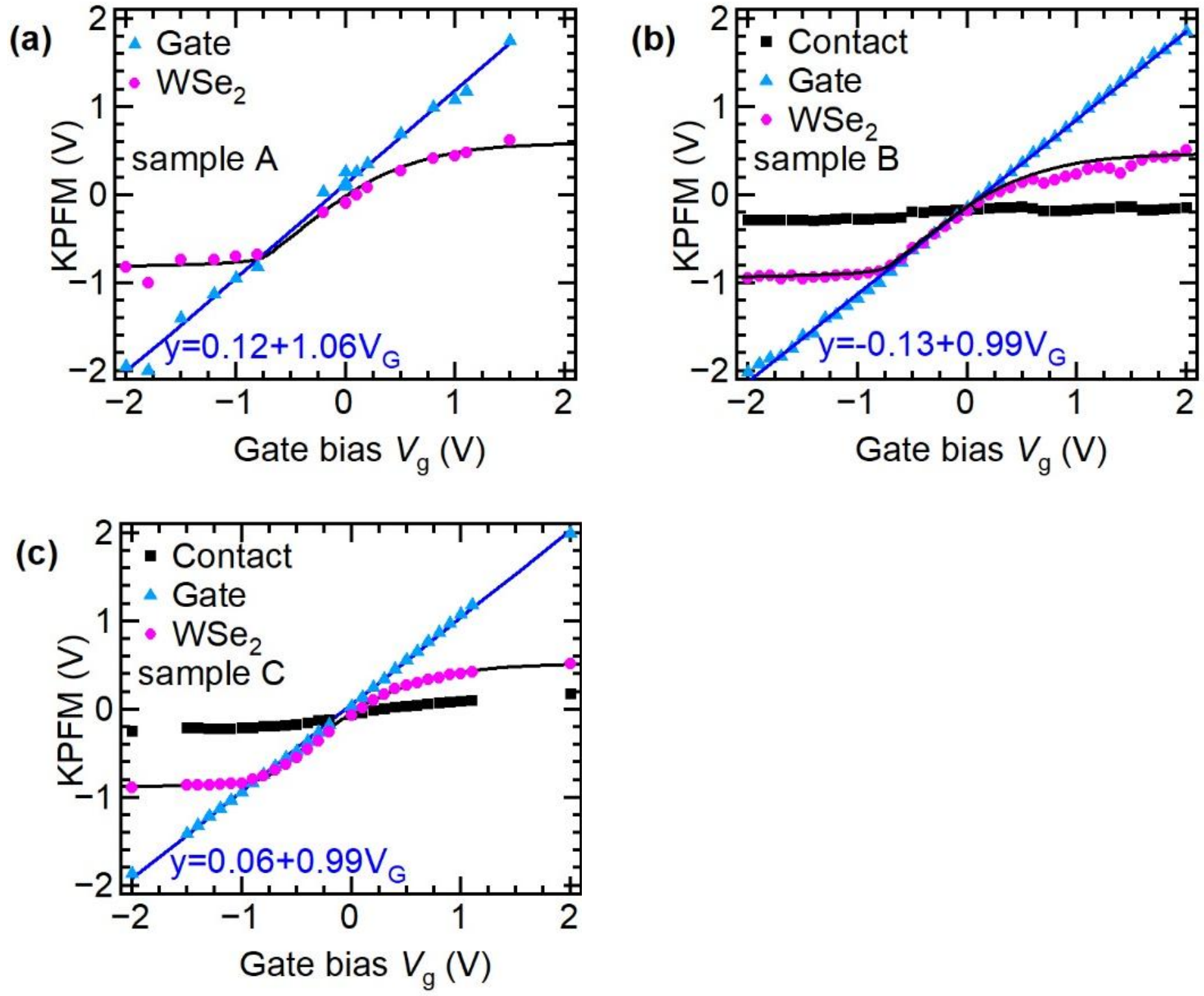

**Figure S6**. Raw data for the three samples with "thin" gate dielectrics. The data over a region that just included the gate under hBN gives a linear response with unitary slope; fit line equation is given in each graph. The mL-Gr contact response varies slightly with gate as expected, consistent with []; differences between samples likely relate to the differences in thickness. (for sample A, the mL-Gr contact was not visible in the image with the other regions). The solid black line is the same model in all three graphs as well as Fig 3b in the main paper, and was originally prepared for sample C. This model does not include interactions between the tip and sample, so in that way it reflects an 'ideal' KPFM measurement. It does include an exponential density of tail states at the conduction band edge as discussed in the paper. (a) sample A, (b) sample B, (c) sample C.

*Alt text: Graphs of KPFM signal versus gate bias for the gate, channel, and contact regions of samples A, B, and C, compared to model results.*

The KPFM signal from the exposed hBN region, and from the Gr contact, provide a test that the electrical connections are good and the KPFM signal seems reasonable. In the exposed hBN region, we expect $V_{\mathrm{KP}}$ to simply track the potential of the underlying graphite flake and thus the gate bias $V_{\mathrm{g}}$, as indeed it does.

Data for sample E, with a 115 nm hBN gate dielectric, was noisier than for the other samples. This isn't surprising, because it has a smaller capacitance, creating a larger voltage response to a change of charge. Lifting the tip compounded the problem, as it decreased tip-sample capacitance, significantly reducing the signal available to the microscope and also (presumably) making the tip response more vulnerable to stray fields in the region. For these reasons, data was not collected at tip heights above 80 nm.

Figure S7 gives the raw data from sample E. The curves shown in the main paper are not shifted from these values. Models use $E_{\mathrm{g}} = 1.1\mathrm{eV}$ or $E_{\mathrm{g}} = 1.2\mathrm{eV}$ as labeled, and this variability in the results may be related to the hysteresis and charge transfer effects described next.

The $W_{\mathrm{T}} = 40\mathrm{nm}$ data corresponds to lowering the average tip height by $20\mathrm{nm}$ during the nap scan. In this case, some hysteresis was visible in the data, as is particularly evident in Fig. S7(a). In all of the datasets for sample E, it's also noticeable that the linear fit line to the exposed hBN/gate region has a larger and more variable intercept compared to the results for thin hBN samples A-C. This isn't surprising considering that the gate capacitance of sample E is about $1/6$ that of samples A-C. Thus, the same amount of stray charge would produce a voltage offset 6x larger in sample E.

As discussed in the paper, this data also seems to show that the parallel plate model overestimates the tip-sample capacitance for larger tip heights. When $W_{\mathrm{T}} = 60\mathrm{nm}$ (see Fig. S6(d)), a parallel plate capacitance model suggests $R = 2.0$, but the modeled results agree better with the data using larger $R$ ($R = 2.7$ is shown). One would expect this effect to be more significant for larger tip heights, and indeed, for $W_{\mathrm{T}} = 80\mathrm{nm}$, for which a parallel plate model gives $R = 2.7$, we found much better agreement using $R = 4.0$. These models are all included in Fig. S6.

This isn't terribly surprising. The tip diameter is about $40\ \mathrm{nm}$, so as the lift height becomes greater than that, a more accurate model is desired. However, the parallel plate model has the advantage of giving a clear intuitive result that can be easily compared to the gate capacitance. And, it gives a reasonable estimate to use in device design, since none of

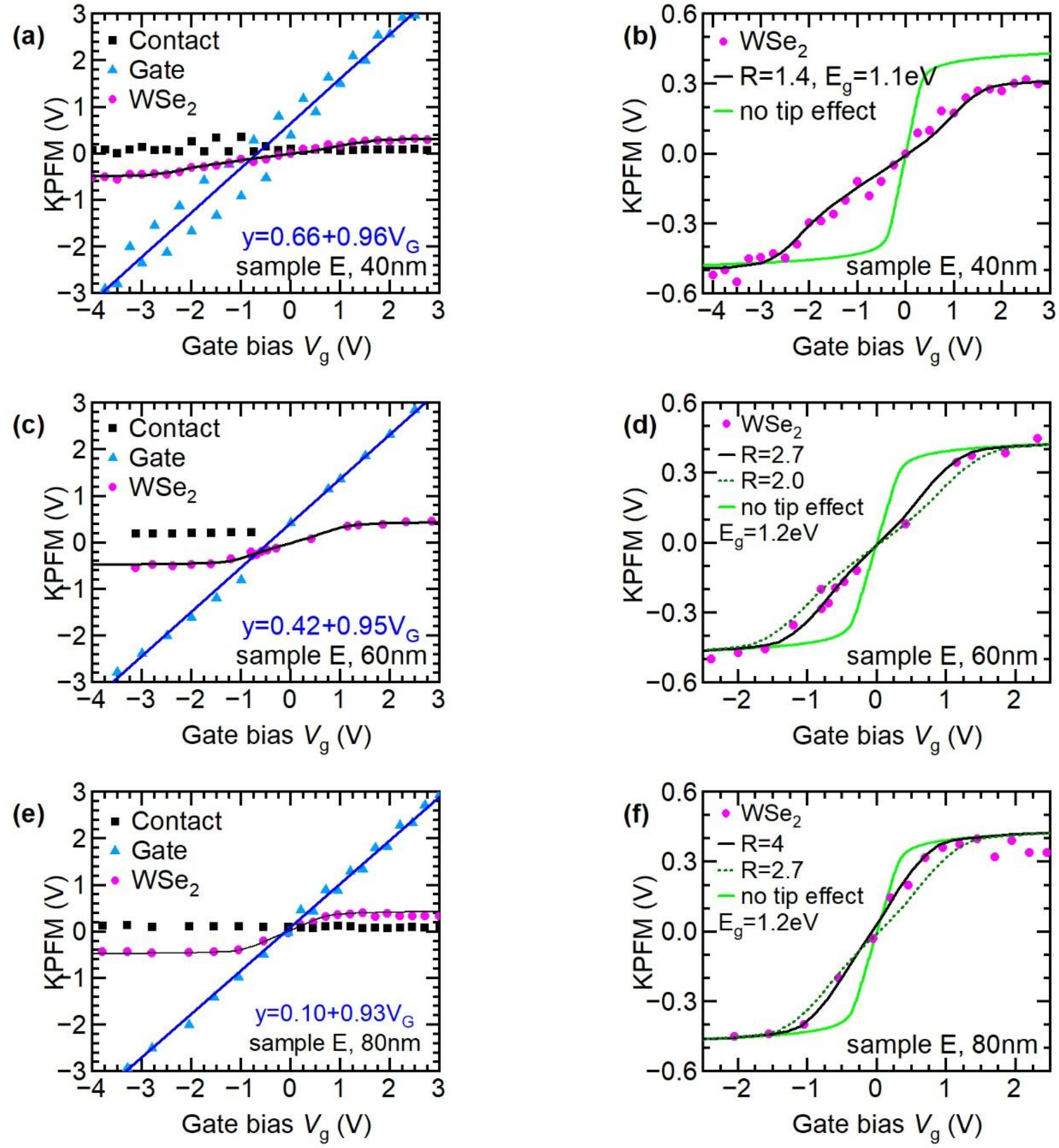

**Figure S7**. FM-KPFM data for sample E, which had $W_{\mathrm{g}} = 115$ nm, at varying tip heights $W_{\mathrm{T}}$ : (**a,b**) $W_{\mathrm{T}} = 40\mathrm{nm}$, (**c,d**) $W_{\mathrm{T}} = 60\mathrm{nm}$, (**e,f**) $W_{\mathrm{T}} = 80\mathrm{nm}$. Parts (**a,c,e**) show signal from three locations: the mL-Gr contact, an exposed hBN/gate region with linear fit line, and the few-layer $WSe_2$ with the model result shown in the main paper (solid black line). In (c), for $W_{\mathrm{T}} = 60$ nm, the mL-Gr contact region was only visible in some datasets. Parts (**b,d,f**) zoom in on the $WSe_2$ response for low gate bias and compare a few different model results as labeled, including the model for no tip effect (green), which uses $E_{\mathrm{g}} = 1.2\mathrm{eV}$.

*Alt text: Graphs of KPFM signal versus gate bias for the gate, channel, and contact regions of sample E, compared to model results.*

these results is highly sensitive to small changes in $R$ (except when $R$ becomes quite small) as can be seen in Fig. S7(d,f).

### Channel potential and gating of monolayer $WSe_2$

The dependence of $V_{\mathrm{CH}}$ on $V_{\mathrm{G}}$, and the charge densities involved, are not entirely intuitive, and so we've included some outputs of our model here. In Fig. S8, when $V_{\mathrm{G}} = 2.0\mathrm{V}$, the value of $V_{\mathrm{CH}}$ seems saturated. However, it's still well below midgap (which is $E_0 = 0.825\mathrm{V}$, the maximum y-axis value). As $V_{\mathrm{G}}$ continues to increase to several tens of V, $V_{\mathrm{CH}}$ does gradually approach $E_0$ (not shown here). This is significantly different from the sharp saturation that occurs in a low-temperature measurement.

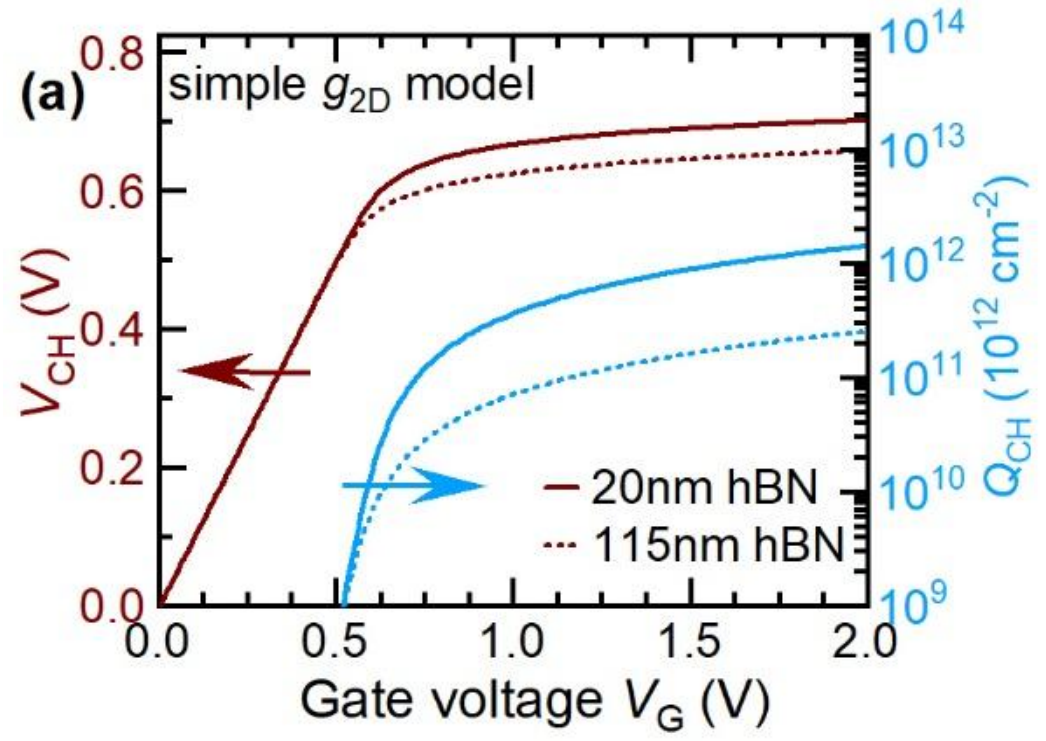

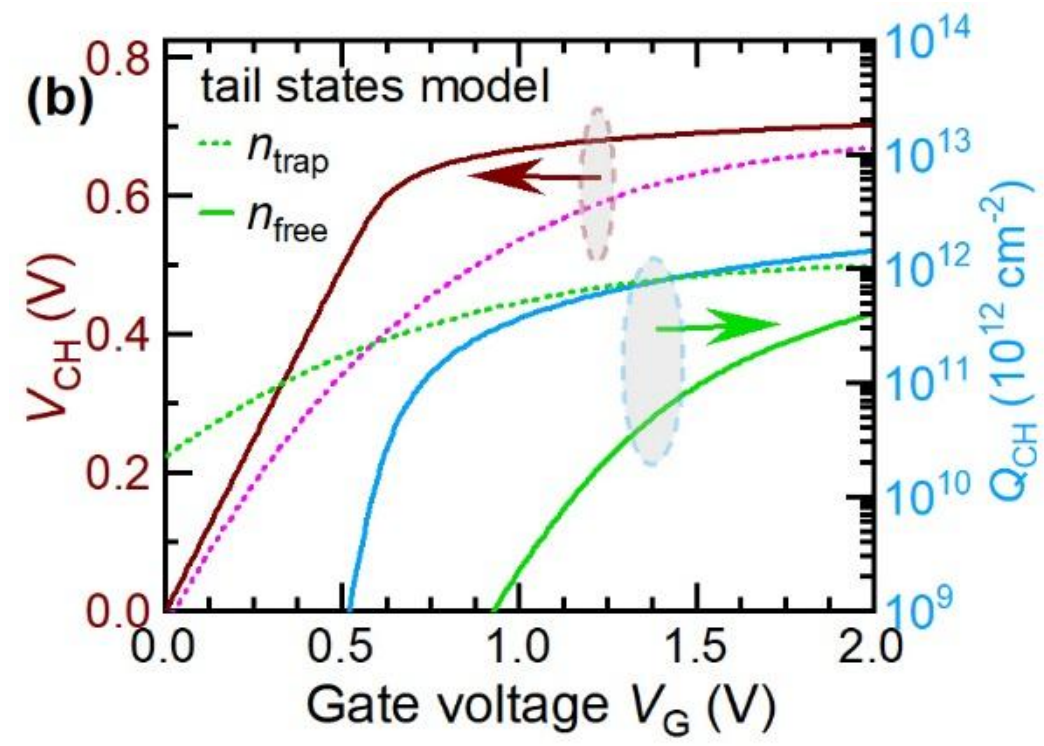

**Figure S8**. Dependence of channel potential $V_{\mathrm{CH}}$ and the associated charge $Q_{\mathrm{CH}}$ on gate voltage. **(a)** Modeled response of monolayer $WSe_2$ ($E_{\mathrm{g}} = 1.65$ eV) gated through a $20$ nm thick hBN gate insulator, similar to sample C (solid lines), compared to gating through a 115 nm thick hBN gate insulator, similar to sample E (dotted lines). For $V_{\mathrm{g}} > 1V$, the decreased gate capacitance of sample E has a relatively minor effect on $V_{\mathrm{CH}}$, but $Q_{\mathrm{CH}}$ is almost an order of magnitude smaller. **(b)** Adding a density of bandtail states to the model, as described in Eq. 6 in the main text. These models all use a $20$ nm thick hBN gate insulator. Graph includes $V_{\mathrm{CH}}$ (pink dotted line), tail state charge density (green dotted line), and free carriers (green solid line). The simple model results from part (a) are also shown for comparison, using the same color scheme.

*Alt text: Graphs showing modeled output of channel potential and channel charge versus gate voltage for typical models.*